\documentclass[twocolumn,showpacs,groupedaddress,
preprintnumbers,amsmath,amssymb,aps,prd]{revtex4}
\def\be{\begin{equation}}
\def\ee{\end{equation}}
\def\bea{\begin{eqnarray}}
\def\eea{\end{eqnarray}}
\def\ba{\begin{array}}
\def\ea{\end{array}}
\def\nn{\nonumber}
\def\p{\partial}

\def\sda{\sin^2\theta}
\def\cda{\cos^2\theta}

\def\Dp{\Delta^{\prime}}
\def\tD{\tilde{\Delta}}
\def\trm{\textrm{Im}}

\begin{document}

\preprint{hep-th/0512351~v4}
\title{Hawking radiation as tunneling from the Kerr and Kerr-Newman black holes}

\author{Qing-Quan Jiang \footnote{E-mail address: jiangqingqua@126.com}}
\affiliation{\centerline{College of Physical Science and Technology,
Central China Normal University, Wuhan, Hubei 430079, People's Republic of China} \\
\centerline{and Institute of Theoretical Physics, China West Normal University,
Nanchong, Sichuan 637002, People's Republic of China}}
\author{Shuang-Qing Wu \footnote{E-mail address: sqwu@phy.ccnu.edu.cn (Corresponding author)}}
\affiliation{\centerline{College of Physical Science and Technology, Central China Normal University,
Wuhan, Hubei 430079, People's Republic of China}}
\author{Xu Cai \footnote{E-mail address: xcai@mail.ccnu.edu.cn}}
\affiliation{\centerline{Institute of Particle Physics, Central China Normal University,
Wuhan, Hubei 430079, People's Republic of China}}
\revised{\today}

\begin{abstract}
Recent work, which treats the Hawking radiation as a semi-classical tunneling process at the horizon of the
Schwarzschild and Reissner-Nordstr\"{o}m spacetimes, indicates that the exact radiant spectrum is no longer
pure thermal after considering the black hole background as dynamical and the conservation of energy. In this
paper, we extend the method to investigate Hawking radiation as massless particles tunneling across the event
horizon of the Kerr black hole and that of charged particles from the Kerr-Newman black hole by taking into
account the energy conservation, the angular momentum conservation, and the electric charge conservation.
Our results show that when self-gravitation is considered, the tunneling rate is related to the change of
Bekenstein-Hawking entropy and the derived emission spectrum deviates from the pure thermal spectrum, but
is consistent with an underlying unitary theory.
\end{abstract}

\pacs{04.70.Dy, 04.62.+v, 03.65.Sq}
\maketitle

\section{Introduction}

The ``no hair'' theorem stated that all information about the collapsing body was lost from the outside
region apart from three conserved quantities: the mass, the angular momentum, and the electric charge. In
other words, this implied that the only stationary rotating black hole solutions of the Einstein-Maxwell
equations in four dimensions are the Kerr-Newman metrics. In the classical theory, the loss of information
was not a serious problem since the information could be thought of as preserved inside the black hole but
just not very accessible. However, taking the quantum effect into consideration, the situation is changed.
With the emission of thermal radiation \cite{SWH}, black holes could lose energy, shrink, and eventually
evaporate away completely. Since the radiation with a precise thermal spectrum carries no information, the
information carried by a physical system falling toward black hole singularity has no way to be recovered
after a black hole has disappeared completely. This is the so-called ``information loss paradox'' \cite{ILP},
which means that pure quantum states (the original matter that forms the black hole) can evolve into mixed
states (the thermal spectrum at infinity). Such an evolution violates the fundamental principles of quantum
theory, as these prescribe a unitary time evolution of basis states. While the information paradox can perhaps
be attributed to the semi-classical nature of the investigations of Hawking radiation, researches in string
theory indeed support the idea that Hawking radiation can be described within a manifestly unitary theory,
however, it still remains a mystery how information is recovered. Although a complete resolution of the
information loss paradox might be within a unitary theory of quantum gravity or string/M-theory, it is
argued that the information could come out if the outgoing radiation were not exactly thermal but had
subtle corrections \cite{ILP}.

On the other hand, the mechanism of black hole radiance remains shrouded in some degree of mystery. In the
original derivation of black hole evaporation, Hawking described the thermal radiation as a quantum tunneling
process \cite{HH} triggered by vacuum fluctuations near the event horizon. According to this scenario, a pair
of particles is spontaneously created just inside the horizon, the positive energy particle then tunnels out
to the infinity, and the negative energy ``partner'' remains behind and effectively lowers the mass of the
black hole. This tunneling picture can be depicted in another manner, that is, a particle/anti-particle pair
is created just outside the horizon, the negative energy particle tunnels into the horizon because the negative
energy orbit exists only inside the horizon, the positive energy ``partner'' is left outside and emerges at
infinity.

In fact, the above viewpoint that regards the radiation as quantum tunneling out from inside the black
hole has been proved very convenient to explore the issue of dynamics. But, actual derivation \cite{HRD}
of Hawking radiation did not proceed in this way at all, most of which based upon quantum field theory on
a fixed background spacetime without considering the fluctuation of the spacetime geometry. Moreover,
there is another fundamental issue that must necessarily be dealt with, namely, the energy conservation.
It seems clear that the background geometry of a radiating black hole should be altered with the loss of
energy, but this dynamical effect is often neglected in formal treatments.

Recently, a program that implemented Hawking radiation as a tunneling process was initiated by Kraus and
Wilczek \cite{KKW} and developed by Parikh and Wilczek \cite{PW}, (this framework shall be referred to as
the Kraus-Parikh-Wilczek's analysis for briefness, see also Ref. \cite{CPM} for a different methodology that
the tunneling picture has been applied.) who carried out a dynamical treatment of black hole radiance in the
static spherically symmetric black hole geometries. More specifically, they considered the effects of a positive
energy matter shell propagating outwards through the horizon of the Schwarzschild and Reissner-Nordstr\"{o}m
black holes, and incorporated the self-gravitation correction of the radiation. In particular, they took into
account the energy conservation and allowed the background geometry to fluctuate in their dynamical description
of the black hole background. In doing so, they allowed the black hole to lose mass while radiating, but maintained
a constant energy for the total system. The emission spectrum that they calculated for the Schwarzschild and
Reissner-Nordstr\"{o}m black holes gives a leading-order correction to the emission rate arising from loss of
mass of the black hole, which corresponds to the energy carried by the radiated quantum. This result displays
that the derived spectrum of black hole radiation is not strictly pure thermal under the consideration of energy
conservation and the unfixed spacetime background, which may be a correct amendment to Hawking radiation spectrum.

Apart from the energy conservation and the particle's self-gravitation are considered, a salient point in
the Kraus-Parikh-Wilczek's analysis is to introduce a coordinate system that is well-behaved at the event
horizon in order to calculate the emission probability. The so-called ``Painlev\'{e}-Gullstrand coordinates''
rediscovered in Ref. \cite{KW} are not only time independent and regular at the horizon, but for which time
reversal is manifestly asymmetric, namely, the coordinates are stationary but not static. Following this approach,
a lot of people \cite{ETR,VM,JW} have investigated Hawking radiation as tunneling from various spherically
symmetric black holes, and the obtained results are very successful to support the Parikh-Wilczek's picture.
Nevertheless, all these investigations are limited to the spherically symmetric black holes and most of them
are confined only to discuss the tunneling process of the uncharged massless particles. There are also some
recent attempts to extend this approach to the case of the stationary axisymmetric geometries \cite{ANVZ,RBH},
however, as far as the treatment is concerned, not all of them are completely satisfactory.

The purpose of the current paper is to present a reasonable extension of the self-gravitation analysis (we will
follow the presentation of \cite{PW}) from spherically symmetric spacetime to the case of a rotating Kerr \cite{Kerr}
black hole. Moreover, we attempt to extend this method to investigate the tunneling radiation of charged particles
from the event horizon of a stationary Kerr-Newman \cite{KN} black hole. In order to do so, one needs to find a
coordinate system that is well-behaved at the event horizon, since it is the key to do a tunneling computation in
the semi-classical framework of \cite{KKW,PW}. Fortunately, several years ago a suitable solution to this problem
was already provided by Chris Doran \cite{NFK}, who presented a faithful generalization of the Painlev\'{e}-Gullstrand
coordinate system to the case of the Kerr black hole. The new form of the Kerr solution found in Ref. \cite{NFK} inherits
most of the attractive properties of the Painlev\'{e}-Gullstrand coordinate system: (1) The metric is regular at the
event horizon; (2) The time direction remains to be a Killing vector besides there exists another Killing vector
$\p_{\phi}$; (3) The time coordinate $t$ registers the local proper time for radially free-falling observers; (4)
The measure on the surfaces of constant-time slices is the same as that of flat spacetime; (5) In addition, it
satisfies Landau's condition of the coordinate clock synchronization \cite{LL}. All these features are very useful
to study the radiation of particles tunneling across the event horizon of a rotating black hole. Their utility will
be demonstrated by how to calculate the tunneling probability in this paper.

Now since we shall adopt the semi-classical method \cite{KKW,PW} with which the gravitation effects need not to be
taken into account in our discussion, it seems that the unique task left is directly to calculate the tunneling rate
and the corrected radiant spectrum of the black holes. Nevertheless, there is still another important issue needed
to be well addressed, namely, the frame-dragging effect of a rotating black hole. In general, because there exists
a frame-dragging effect of the coordinate system in the stationary rotating spacetime, the matter field in the
ergosphere near the horizon must be dragged by the gravitational field with an azimuthal angular velocity also,
so a legitimate physical picture should be described in the dragging coordinate system. In addition, due to the
presence of rotation, the event horizon does not coincide with the infinite red-shift surface in both forms of
the original Kerr \cite{Kerr} solution and that presented by Doran \cite{NFK}, the geometrical optical limit cannot
be used there since the Kraus-Parikh-Wilczek's analysis is essentially akin to a WKB (`$s$-wave') approximation.
So the Painlev\'{e}-Kerr metric introduced by Doran \cite{NFK} is still inconvenient for us to depict the tunneling
process that takes place at the event horizon. Apparently, this superficial difficulty can be easily overcome by
further performing a dragging coordinate transformation which makes the event horizon coincide with the infinite
red-shift surface so that the WKB approximation can be applied now. The final Painlev\'{e}-Kerr coordinate system
at which we arrived in a dragging coordinate system retains all the nice features mentioned above, therefore it
is very convenient for us to investigate the tunneling process of a rotating black hole.

After equipped with these insights, we are readily to investigate Hawking radiation of massless particles as a
tunneling process across the event horizon of the Kerr black hole and that of charged particles from the Kerr-Newman
black hole by taking into account the energy conservation, the angular momentum conservation, and the electric charge
conservation. The picture adopted in our discussion is: a (charged) particle does tunnel out of a rotating (charged)
black hole, the tunneling barrier is created by the self-gravitation among the outgoing particle. If the total energy
and total angular momentum as well as electric total charge must be conserved, the outgoing particle must tunnel out a
radial barrier to an observer resting in the dragging coordinate system. With the loss of energy, the angular momentum,
and the electric charge, the black hole will shrink its size, its dragging velocity and electric potential will change
also. That is to say, the geometry must be dynamic. This picture enables us to compute the tunneling rate and the
radiant spectrum of a rotating Kerr and Kerr-Newman black hole. Our results show that when self-gravitation is
considered, the tunneling probability is related to the change of Bekenstein-Hawking entropy and the derived
emission spectrum deviates from the pure thermal spectrum, but is consistent with an underlying unitary theory.

Our paper is outlined as follows. In Sec. \ref{NFKDC}, we begin with by briefly reviewing the extension of the
Painlev\'{e}-Gullstrand coordinate system to the Kerr black hole case and show that the new form \cite{NFK}
of the Kerr solution has many of the desired properties to do a semi-classical tunneling calculation of the
emission rate from which the corrected spectrum can be directly extrapolated. We then introduce the dragging
coordinate system in the Painlev\'{e}-Kerr spacetime and present the radial geodesic equation of uncharged
particles. In Sec. \ref{KerrTR}, we investigate Hawking radiation as tunneling from the event horizon of the
Kerr black hole and compute the tunneling rate. In Sec. \ref{KNTR}, we extend this analysis to Kerr-Newman
spacetime and calculate the emission probability of charged particles. Section \ref{CoRe} ends with some brief
remarks.

\section{New form of the Kerr solution and dragging coordinate system}
\label{NFKDC}

Recall that in the tunneling framework advocated by Parikh and Wilczek's \cite{PW}, a convenient trick is to use
stationary coordinates that are manifestly asymmetric under time reversal. In the case of a Schwarzschild black
hole, it is most convenient to recast the metric into the form of Painlev\'{e}-Gullstrand coordinates
\be
ds^2 = dt^2 -\Big(\sqrt{\frac{2M}{r}}dt +dr\Big)^2 -r^2(d\theta^2 +\sda d\phi^2) \, .
\label{SPcs}
\ee

The Painlev\'{e}-Gullstrand coordinate system (\ref{SPcs}) has a number of nice features \cite{NFK}, many of which extend to
the Kerr case. The solution is well-behaved, without a singularity at the horizon, so can be employed safely to
analyze physical processes near the event horizon, and indeed inside it. A second useful feature is that the time
coordinate $t$ coincides with the local proper time of observers free-falling along radial trajectories starting
from rest at infinity. Because it has many of the properties of a global, Newtonian time, physics as seen by these
observers is almost entirely Newtonian, making it a very powerful one for studying across-horizon physics. Another
useful property of this metric is that the measure on surfaces of constant-time slices is the same as that of flat
Euclidean spacetime. A further feature of the metric is that an observer at infinity does not make any distinction
between the Painlev\'{e}-Gullstrand coordinates and the static Schwarzschild coordinates, the two time coordinates coincides
with each other there. Finally, the line element satisfies the Landau's condition of coordinate clock synchronization
\cite{LL}, making it very important to discuss the tunneling process because particle tunneling through a barrier
is an instantaneous process in the sense of quantum mechanics.

Extending to the case of a stationary rotating black hole, it is naturally expected to find an analogue of the
Painlev\'{e}-Gullstrand coordinates for the Kerr solution. Such an attempt might fail due to the presence of angular momentum.
However, Doran \cite{NFK} indeed achieved a suitable generalization by realizing that the key is to look for a
convenient set of reference observers which generalizes the idea of a family of free-falling observers on radial
trajectories, since it is only the local properties of time $t$ that make it so convenient for describing the
physics of the solution. Starting from the advanced Eddington-Finkelstein coordinate formalism of the Kerr metric
and performing a coordinate transformation, Doran reached to a new form of the Kerr metric as follows \cite{NFK}
\bea
ds^2 &=& dt^2 -\Big[\sqrt{\frac{2Mr}{\Sigma}}\big(dt -a\sda d\phi\big)
 +\sqrt{\frac{\Sigma}{r^2 +a^2}}dr\Big]^2 \nn \\
&&~  -\Sigma d\theta^2 -(r^2 +a^2)\sda d\phi^2 \, ,
\label{NewK}
\eea
where $\Sigma = r^2 +a^2\cda$, and $\Delta = r^2 +a^2 -2Mr$ (see below), in which $M$ is the mass, the specific
angular momentum $a = J/M$ is kept as a constant thought this paper.

Unlike Doran did in Ref. \cite{NFK}, instead we shall demonstrate that via a suitable coordinate transformation
the metric (\ref{NewK}) can be directly obtained from the usual form of the Kerr solution which can be expressed
as \cite{Kerr}
\bea
ds^2 &=& d\bar{t}^2 -\frac{2Mr}{\Sigma}\Big(d\bar{t} -a\sda d\bar{\phi}\Big)^2
-\frac{\Sigma}{\Delta}dr^2 \nn \\
&& -\Sigma d\theta^2 -(r^2 +a^2)\sda d\bar{\phi}^2 \, .
\eea
In order to do so, it is instructive to note that the metric (\ref{SPcs}) can be obtained from the original
Schwarzschild solution
\be
ds^2 = \Big(1 -\frac{2M}{r}\Big) d\bar{t}^2 -\frac{dr^2}{1 -2M/r}
 -r^2\big(d\theta^2 +\sda d\bar{\phi}^2\big) \, ,
\ee
by a coordinate transformation of the Painlev\'{e}-type
\be
d\bar{t} = dt -\frac{\sqrt{2Mr}}{r -2M}dr \, .
\ee
So a natural extension of this transformation in the rotating case should be written as
\bea
d\bar{t} &=& dt -\frac{\sqrt{2Mr(r^2 +a^2)}}{\Delta}dr \, , \nn \\
d\bar{\phi} &=& d\phi -\frac{a}{\Delta}\sqrt{\frac{2Mr}{r^2 +a^2}}dr \, .
\eea
A direct computation can check this indeed is the case.

The new form (\ref{NewK}) of the Kerr metric directly generalizes the Painlev\'{e}-Gullstrand metric (\ref{SPcs}),
replacing $\sqrt{2M/r}$ with $\sqrt{2Mr/\Sigma}$, and introducing a rotational component. This coordinate system
faithfully inherits a number of nice characters of the Painlev\'{e}-Gullstrand line element: (1) The metric is
well-behaved at the event horizon; (2) There exist two commuting Killing vectors $\p_t$ and $\p_{\phi}$; (3) The
time coordinate $t$ represents the local proper time for radially free-falling observers; (4) The hypersurfaces
of constant-time slices are just flat Euclidean space in the oblate spheroidal coordinates; (5) In addition, it
satisfies Landau's condition of the coordinate clock synchronization \cite{LL}. As such, we shall refer it to as
the Painlev\'{e}-Kerr coordinate system. Since these coordinates comply with the perspective of a free-falling
observer, who is expected to experience nothing out of the ordinary upon passing through the horizon, it is
well-suited for studying processes near the event horizon. The new form of the Kerr solution has already proved
to be very powerful in numerical simulation in astrophysics, black hole physics and quantum mechanics in curved
spaces (\cite{NFK} and references therein). In this paper, we shall further exploit its another application in
black hole phenomena due to its distinguishing feature of horizon regularity, namely, Hawking radiation as
tunneling from the horizon.

However, due to the inclusion of rotation, the Painlev\'{e}-Kerr metric (\ref{NewK}) is still inconvenient for our
study of the tunneling process at the event horizon. The reasons come from two aspects. In the one hand, because
the event horizon $r_+ = M +\sqrt{M^2 -a^2}$ does not coincide with the infinite red-shift surface $r_{TLS} = M
+\sqrt{M^2 -a^2\cda}$ in both forms of the original Kerr solution and the Painlev\'{e}-Kerr metric, the geometrical
optical limit cannot be applied. On the other hand, since there exists a frame-dragging effect in the stationary
rotating spacetime, the matter field in the ergosphere near the horizon must be dragged by the gravitational field
also, so a reasonable physical picture should be depicted in the dragging coordinate system. Obviously, this hints
that we must continue to transform the metric into a dragging coordinate system. Carrying out a dragging coordinate
transformation $d\phi = \Omega~dt$ with
\be
\Omega = \frac{d\phi}{dt} = -\frac{g_{t\phi}}{g_{\phi\phi}}
= \frac{a(r^2 +a^2 -\Delta)}{(r^2 +a^2)^2 -\Delta a^2\sda} \, ,
\label{AV}
\ee
yields the new line element, which shall be called as the dragged Painlev\'{e}-Kerr metric
\bea
d\hat{s}^2 &=& \frac{\Delta\Sigma}{(r^2 +a^2)^2 -\Delta a^2\sda}dt^2 -\frac{\Sigma}{r^2 +a^2}dr^2 \nn \\
&& -2\frac{\sqrt{2Mr(r^2 +a^2)}\Sigma}{(r^2 +a^2)^2 -\Delta a^2\sda} dtdr -\Sigma d\theta^2 \, .
\label{DPK}
\eea

In fact, the line element (\ref{DPK}) represents a 3-dimensional hypersurface in the 4-dimensional spacetime, with no
coordinate singularity at the horizon. Along with the above-mentioned properties of the Painlev\'{e}-Kerr coordinate
system, the event horizon and the infinite red-shift surface coincide with each other in the dragged Painlev\'{e}-Kerr
coordinate system so that the WKB approximation can be used now. These attractive features are very advantageous for
us to discuss Hawking radiation via tunneling and to do an explicit computation of the tunneling probability at the
event horizon.

In the subsequent section, we shall investigate the tunneling behavior of massless particles from the horizon. For
this purpose, let us first evaluate the radial, null geodesics. Since the tunneling processes take place near the
event horizon, we may consider a particle tunneling across the event horizon as an ellipsoid shell and think that
the particle should still be an ellipsoid shell during the tunneling process, i.e., the particle does not have motion
in the $\theta$-direction. Therefore, under these assumptions ($d\hat{s}^2 = 0 = d\theta$), the radial, null geodesics
followed by massless particles are
\bea
\dot{r} &=& \frac{dr}{dt} = \frac{(r^2 +a^2)^2}{(r^2 +a^2)^2 -\Delta a^2\sda} \nn \\
&& \times \Bigg[-\sqrt{1 -\frac{\Delta}{r^2 +a^2}} \pm
\sqrt{1 -\frac{\Delta^2a^2\sda}{(r^2 +a^2)^3}}\Bigg] \, , \quad
\label{RNG}
\eea
where the upper (lower) sign can be identified with the outgoing (incoming) radial motion, under the implicit
assumption that time $t$ increases towards the future. In other words, the plus sign corresponds to an outgoing
geodesic and the minus sign corresponds to an ingoing geodesic, respectively.

\section{Tunneling process of uncharged particles from Kerr black holes}
\label{KerrTR}

Now we turn to discuss Hawking radiation of uncharged particles as a semi-classical tunneling process across the
barrier which is created just by the outgoing particle itself. We adopt the picture of a pair of virtual particles
spontaneously created just inside the horizon. The positive energy virtual particle can tunnel out and materialize
as a real particle escaping classically to infinity, its negative energy partner is absorbed by the black hole,
resulting in a decrease in the mass and angular momentum of the black hole. In our discussion, we consider the
particle as an ellipsoid shell of energy $\omega$ and angular momentum $a\omega$. If the particle's self-gravitation
is taken into account, Eqs. (\ref{AV}-\ref{RNG}) should be modified. To guarantee the conservation of energy and
angular momentum, we fix the total mass and total angular momentum of the spacetime but allow the hole's mass and
angular momentum to fluctuate. When a particle of energy $\omega$ is radiated from the event horizon, the mass and
angular momentum of the black hole will be reduced to $M -\omega$ and $(M -\omega)a$, respectively. Then we should
replace $M$ with $M -\omega$ in Eqs. (\ref{AV}-\ref{RNG}) in order to describe the moving of the shell. In particular,
the particle will move along the modified null geodesic in the radial direction
\be
\dot{r} = \frac{\tD}{r^2 +a^2}
\Bigg[\sqrt{1 -\frac{\tD}{r^2 +a^2}} \pm \sqrt{1 -\frac{{\tD}^2a^2\sda}{(r^2 +a^2)^3}}\Bigg]^{-1}\, ,
\label{MG}
\ee
where $\tD = r^2 +a^2 -2(M -\omega)r$ is the horizon equation after the emission of the particle with energy $\omega$.

Since the event horizon coincides with the infinite red-shift surface in the dragged Painlev\'{e}-Kerr coordinate
system, so the geometrical optical limit become an especially reliable approximation and the semi-classical WKB
(`$s$-wave') approximation can be used. By means of WKB approximation, the tunneling probability for an outgoing
positive energy particle can be expressed in terms of the imaginary part of the action as
\be
\Gamma \sim e^{-2 \trm~ S} \, .
\ee
At this point, it should be noticed that the coordinate $\phi$ does not appear in the dragged Painlev\'{e}-Kerr
metric (\ref{DPK}). That is, $\phi$ is an ignorable coordinate in the Lagrangian function. To eliminate this degree
of freedom completely, the imaginary part of the action should be written as [by using $dt = d\phi/\dot{\phi} =
dr/\dot{r}$.]
\bea
\trm~ S &=& \trm \int_{t_i}^{t_f}\big(P_r\dot{r} -P_{\phi}\dot{\phi}\big)dt \nn \\
&=& \trm\int\limits_{r_i}^{r_f}\Bigg[\int\limits_{(0, ~0)}^{(P_r, P_{\phi})}
 \Big(\dot{r}~dP_r^{\prime} -\dot{\phi}~dP_{\phi}^{\prime}\Big)\Bigg]\frac{dr}{\dot{r}} \, ,
 \label{IPA}
\eea
where $P_r$ and $P_{\phi}$ are two canonical momenta conjugate to $r$ and $\phi$, respectively. $r_i = r_+ = M
+\sqrt{M^2 -a^2}$ and $r_f = M -\omega +\sqrt{(M-\omega)^2 -a^2}$ are the locations of the event horizon before
and after a particle tunnels out, they are just inside and outside the barrier through which the particle tunnels.

To proceed with an explicit calculation, we now remove the momentum in favor of energy by applying the Hamilton's
equations
\bea
\dot{r} &=& \frac{dH}{dP_r}\Big|_{(r; \phi, P_{\phi})}
 = \frac{d(M -\omega)}{dP_r} \, , \nn \\
\dot{\phi} &=& \frac{dH}{dP_{\phi}}\Big|_{(\phi; r, P_r)}
 = a\tilde{\Omega}\frac{d(M -\omega)}{dP_{\phi}} \, ,
\label{HE}
\eea
where $dH_{(\phi; r, P_r)} = \tilde{\Omega}dJ = a\tilde{\Omega}d(M -\omega)$, which represents the energy change
of the black hole because of the loss of the angular momentum when a particle tunnels out, and the dragging angular
velocity is given by
\bea
\tilde{\Omega} = \frac{a(r^2 +a^2 -\tD)}{(r^2 +a^2)^2 -\tD a^2\sda} \, . \nn
\eea
Substituting Eqs. (\ref{MG}) and (\ref{HE}) into Eq. (\ref{IPA}), and noting that we must choose the positive
sign in Eq. (\ref{MG}) as the particle is propagating from inside to outside the event horizon, then we have
\bea
\trm~ S &=& \trm\int_{r_i}^{r_f}\int_M^{M -\omega}
 \Big[(1 -a\Omega^{\prime})d(M -\omega^{\prime})\Big]\frac{dr}{\dot{r}} \nn \\
 &=& \trm\int_M^{M -\omega}\int_{r_i}^{r_f} \frac{r^2(r^2 +a^2)
 +\Dp a^2\cda}{(r^2 +a^2)^2 -\Dp a^2\sda} \nn \\
 && \times \Big(\frac{r^2 +a^2}{\Dp}\Big)\Bigg[\sqrt{1 -\frac{\Dp}{r^2 +a^2}} \nn \\
&& +\sqrt{1 -\frac{{\Dp}^2a^2\sda}{(r^2 +a^2)^3}}\Bigg] dr d(M -\omega^{\prime}) \, ,
\label{IA}
\eea
where
\bea
&& \Dp =  r^2 +a^2 -2(M -\omega^{\prime})r = (r -r_+^{\prime})(r -r_-^{\prime}) \, , \nn \\
&& r_{\pm}^{\prime} = M -\omega^{\prime} \pm \sqrt{(M -\omega^{\prime})^2 -a^2} \, . \nn
\eea
We see that $r = r_+^{\prime}$ is a single pole in Eq. (\ref{IA}). The integral can be evaluated by deforming
the contour around the pole, so as to ensure that positive energy solution decay in time. In this way, we finish
the $r$ integral and get
\begin{widetext}
\be
\trm~ S = -2\pi\int_M^{M -\omega} \frac{(M -\omega^{\prime})^2 +(M -\omega^{\prime})
 \sqrt{(M -\omega^{\prime})^2 -a^2} -a^2/2}{\sqrt{(M -\omega^{\prime})^2 -a^2}} d(M -\omega^{\prime}) \, .
\ee
\end{widetext}
Completing the integration finally yields
\bea
\trm~ S &=& \pi\Big[M^2 -(M -\omega)^2 +M\sqrt{M^2 -a^2} \nn \\
&&\quad -(M -\omega)\sqrt{(M -\omega)^2 -a^2}\Big] \, .
\eea
In terms of the entropy expression $S_{BH} = \pi(r_+^2 +a^2)$, the tunneling rate is then expressible as \cite{KKW}
\be
\Gamma \sim e^{-2\trm~ S} = e^{\Delta S_{BH}} \, ,
\ee
where $\Delta S_{BH} = S_{BH}(M -\omega) -S_{BH}(M)$ is the difference of Bekenstein-Hawking entropies of the
Kerr black hole before and after the emission of the particle. The derived emission spectrum actually deviates
from pure thermality.

To conclude this section, we find that in order to properly extend the semi-classical tunneling formalism
\cite{KKW,PW} to the case of a Kerr black hole, we must adopt the Painlev\'{e}-Kerr metric which neatly
generalizes the Painlev\'{e}-Gullstrand line element. Nevertheless, we must further transform it to the
dragged Painlev\'{e}-Kerr coordinate system so that an explicit tunneling analysis can be made. Moreover,
when the energy conservation and the angular momentum conservation as well as the particle's self-gravitation
are taken into account, the tunneling rate is related to the change of black hole entropy during the process
of the particle's emission and the radiant spectrum is not precisely thermal. Furthermore, it should be pointed
out that our discussion made in this section can be directly generalized to deal with the Hawking radiation of
uncharged particles tunneling from the Kerr-Newman black hole with a simple replacement of $2Mr$ by $2Mr -Q^2$.
The result is the same as that obtained above, generalizing those given in Refs. \cite{KKW,PW}. In the next
section, we will focus on a semi-classical treatment of the tunneling characters of charged massive particles
from a charged Kerr black hole.

\section{Radiation of charged particles as tunneling from the Kerr-Newman black hole}
\label{KNTR}

As mentioned in the last section, the analysis of uncharged massless particles tunneling from a Kerr-Newman black
hole completely parallels to the case that made for a Kerr black hole. In this section, we shall investigate the
tunneling behavior of a charged massive particle and calculate its emission rate from a charged rotating black
hole. It should be noted that one must overcome two additional difficulties. The first is that one has to decide
the equation of motion of a charged test particle since the radial, null geodesics is only applicable to describe
the tunneling behavior of the uncharged radiation from the event horizon. Different from the null geodesics of
an uncharged particle, the trajectory followed by a charged massive particle is not light-like, but subject to
Lorentz forces. Here we will decide it approximately by the phase velocity. The second is how to take into account
the effect of the electro-magnetic field when the charged particle tunnels out from the event horizon. Apart from
the conservation of energy and angular momentum, the electric charge conservation must be considered also. In the
following discussion, we shall adopt a slightly modified tunneling picture, that is, we consider the charged massive
particle as a charged conducting ellipsoid shell carrying energy $\omega$, angular momentum $a\omega$, and electric
charge $q$. To account for the effect of the electro-magnetic field, we shall consider a matter-gravity system that
consists of the black hole and the electro-magnetic field outside the hole. Taking into account the particle's
self-gravitation, the conservation of energy and angular momentum as well as electric charge, we must fix the
total mass, total angular momentum, and total electric charge of the spacetime but allow those of the black hole
to vary also.

Before seeking for the equation of motion of a charged massive particle, we must first deal with how the
electro-magnetic vector potential changes under two steps of coordinate transformations. To this end, recall
that the Kerr-Newman black hole solution \cite{KN} can be expressed in the Boyer-Lindquist coordinate system as
\bea
ds^2 &=& d\bar{t}^2 -\frac{2Mr -Q^2}{\Sigma}\Big(d\bar{t} -a\sda d\bar{\phi}\Big)^2
-\frac{\Sigma}{\Delta}dr^2 \nn \\
&& -\Sigma d\theta^2 -(r^2 +a^2)\sda d\bar{\phi}^2 \, , \label{KNM} \\
\mathcal{A} &=& \frac{Qr}{\Sigma}\Big(d\bar{t} -a\sda d\bar{\phi}\Big) \, ,
\label{EMP}
\eea
where $\Sigma = r^2 +a^2\cda$, and $\Delta = r^2 +a^2 +Q^2 - 2Mr$, in which the parameters $M$, $Q$, and $J = Ma$
are the mass, the electric charge, and the angular momentum of the black hole, respectively.

A generalized Painlev\'{e}-type coordinate transformation
\bea
d\bar{t} &=& dt -\frac{\sqrt{(2Mr -Q^2)(r^2 +a^2)}}{\Delta}dr \, , \nn \\
d\bar{\phi} &=& d\phi -\frac{a}{\Delta}\sqrt{\frac{2Mr -Q^2}{r^2 +a^2}}dr \, ,
\eea
sends the metric (\ref{KNM}) and the vector potential (\ref{EMP}) to
\bea
ds^2 &=& dt^2 -\Sigma d\theta^2 -\Big[\sqrt{\frac{2Mr -Q^2}{\Sigma}}\big(dt -a\sda d\phi\big) \nn \\
&&  +\sqrt{\frac{\Sigma}{r^2 +a^2}}dr\Big]^2 -(r^2 +a^2)\sda d\phi^2 \, , \\
\mathcal{A} &=& \frac{Qr}{\Sigma}\big(dt -a\sda d \phi\big) \, .
\eea
Here, the form of vector potential remains unchanged up to a gauge transformation. To make the event horizon
coincide with the infinite red-shift surface, we further introduce the dragging coordinate transformation
\be
d\phi = \frac{a(r^2 +a^2 -\Delta)}{(r^2 +a^2)^2 -\Delta a^2\sda}dt \, ,
\ee
under which the desired 3-dimensional dragged Painlev\'{e}-Kerr-Newman line element and the relevant electro-magnetic
vector potential can be obtained as follows
\bea
d\hat{s}^2 &=& \frac{\Delta\Sigma}{(r^2 +a^2)^2 -\Delta a^2\sda}dt^2 -\frac{\Sigma}{r^2 +a^2}dr^2 \nn \label{DPKN} \\
&& -2\frac{\sqrt{(2Mr -Q^2)(r^2 +a^2)}\Sigma}{(r^2 +a^2)^2 -\Delta a^2\sda} dtdr -\Sigma d\theta^2 \, , \quad \\
\hat{\mathcal{A}} &=& \frac{Qr(r^2 +a^2)}{(r^2 +a^2)^2 -\Delta a^2\sda}dt = A_tdt \, .
\label{DEM}
\eea

As before, the radial, null geodesics are given by Eq. (\ref{RNG}) with $2Mr$ replaced by $2Mr -Q^2$. But we are
considering the tunneling process of charged particles from a rotating charged black hole, the trajectory followed
by a charged test particle is not light-like, it does not follow the radially light-like geodesics when it tunnels
across the horizon. Because the calculations are more involved by the fact that the trajectory is now also subject
to Lorentz forces, for the sake of simplicity, here it is  approximately determined by the phase velocity. According
to de Broglie's hypothesis and the definition of the phase (group) velocity, the outgoing particle that can be
considered as a massive shell corresponds to a kind of `$s$-wave' whose phase velocity $v_p$ and group velocity
$v_g$ have the following relationship
\be
v_p = \frac{1}{2}v_g \, ; \qquad\quad v_p = \frac{dr}{dt} \, , \quad v_g = \frac{dr_c}{dt} \, ,
\ee
where $r_c$ denotes the radial position of the particle.

Since the tunneling process across the barrier is an instantaneous effect, there are two events that take place
simultaneously in different places during the process. One is the particle tunneling into the barrier, another
is the particle tunneling out the barrier. Because the dragged Painlev\'{e}-Kerr-Newman metric (\ref{DPKN})
satisfies Landau's condition of the coordinate clock synchronization \cite{LL}, the coordinate time difference
of these two events is
\be
dt = -\frac{g_{tr}}{g_{tt}}dr_c \, , \qquad (d\theta = 0).
\ee
By the definition of the group velocity, we have
\be
v_g = \frac{dr_c}{dt} = -\frac{g_{tt}}{g_{tr}} = \frac{\Delta}{\sqrt{(2Mr -Q^2)(r^2 +a^2)}} \, ,
\ee
therefore the phase velocity is
\be
\dot{r} = \frac{dr}{dt} = -\frac{g_{tt}}{2g_{tr}}
= \frac{\Delta}{2(r^2 +a^2)}\Big/\sqrt{1 -\frac{\Delta}{r^2 +a^2}} \, .
\label{VP}
\ee
To include the particle's self-interaction effect after the charged particle emission, the mass and charge
parameters in Eqs. (\ref{DPKN}, \ref{DEM}) and (\ref{VP}) should be replaced with $M\to M -\omega$ and $Q\to
Q -q$, when a charged test particle of energy $\omega$ and electric charge $q$ tunnels out. Based on a similar
discussion to that made in the last section, it is obvious that we must accordingly modify the radial trajectory
of the charged massive particle to account for the particle's self-gravitation, which is described by
\be
\dot{r} = \frac{\tD}{2(r^2 +a^2)}\Big/\sqrt{1 -\frac{\tD}{r^2 +a^2}} \, ,
\label{CMT}
\ee
where $\tD = r^2 +a^2 -2(M -\omega)r +(Q -q)^2$ is the horizon equation after the emission of the particle with
energy $\omega$ and electric charge $q$.

When we investigate the tunneling process of a charged particle, the effect of the electro-magnetic field should
be taken into account. So we must consider the matter-gravity system that consists of the black hole and the
electro-magnetic field outside the black hole. As the Lagrangian function of the electro-magnetic field corresponding
to the generalized coordinates described by $A_{\mu}$ is $-(1/4)F_{\mu\nu}F^{\mu\nu}$, we can find that the generalized
coordinate $A_t$ in Eq. (\ref{DEM}) is an ignorable coordinate. In addition, the coordinate $\phi$ is also a cyclic
one. In order to eliminate these two degrees of freedom completely, the action of the charged massive particle
should be written as
\bea
S &=& \int_{t_i}^{t_f}\big(P_r\dot{r} -P_{\phi}\dot{\phi} -P_{A_t}\dot{A_t}\big)dt \nn \\
&=& \int\limits_{r_i}^{r_f}\Bigg[\hspace{-0.1cm}\int\limits_{(0, ~0, ~0)}^{(P_r, P_{\phi}, P_{A_t})}
\hspace{-0.3cm} \Big(\dot{r}~dP_r^{\prime} -\dot{\phi}~dP_{\phi}^{\prime}
-\dot{A_t}~dP_{A_t}^{\prime}\Big)\Bigg]\frac{dr}{\dot{r}} \, , \quad~~
\label{CPA}
\eea
where the canonical momenta $\{P_r, P_{\phi}, P_{A_t}\}$ conjugate to the coordinates $\{r, \phi, A_t\}$, $r_i =
M +\sqrt{M^2 -Q^2 -a^2}$ and $r_f = M -\omega +\sqrt{(M -\omega)^2 -(Q -q)^2 -a^2}$ are the locations of the event
horizon before and after the charged particle emission.

According to the Hamilton's equations, we have
\bea
\dot{r} &=& \frac{dH}{dP_r}\Big|_{(r; \phi, P_{\phi}, A_t, P_{A_t})}
 = \frac{d(M -\omega)}{dP_r} \, , \nn \\
\dot{\phi} &=& \frac{dH}{dP_{\phi}}\Big|_{(\phi; r, P_r, A_t, P_{A_t})}
 = a\Omega\frac{d(M -\omega)}{dP_{\phi}} \, , \nn \\
\dot{A_t} &=& \frac{dH}{dP_{A_t}}\Big|_{(A_t; r, P_r, \phi, P_{\phi})}
 = \Phi\frac{d(Q -q)}{dP_{A_t}} \, ,
\label{HES}
\eea
where the dragging angular velocity and the electric potential in the dragging coordinate system are given by
\bea
&& \Omega = \frac{a(r^2 +a^2 -\tD)}{(r^2 +a^2)^2 -\tD a^2\sda}\, , \nn \\
&& \Phi = \frac{(Q -q)r(r^2 +a^2)}{(r^2 +a^2)^2 -\tD a^2\sda} \, . \nn
\eea
We would like to emphasize that, by keeping the mass $M$ and the electric charge $Q$ fixed, the conservation
of energy and angular momentum as well as electric charge will be enforced in a natural fashion.
Substituting Eqs. (\ref{CMT}) and (\ref{HES}) into Eq. (\ref{CPA}), and switching the order of integration
yield the imaginary part of the action
\begin{widetext}
\bea
\trm~ S &=& \trm \int_{r_i}^{r_f}\int_{(M, ~Q)}^{(M -\omega, ~Q -q)}
\Big[(1 -a\Omega^{\prime})d(M -\omega^{\prime}) -\Phi^{\prime}d(Q -q^{\prime})\Big] \frac{dr}{\dot{r}} \nn \\
&=& \trm\int_{(M, ~Q)}^{(M -\omega, ~Q -q)}\int_{r_i}^{r_f}
\frac{2(r^2 +a^2)}{\Delta^{\prime}} \sqrt{1 -\frac{\Delta^{\prime}}{r^2 +a^2}}
\Bigg[\frac{r^2(r^2 +a^2) +\Dp a^2\cda}{(r^2 +a^2)^2 -\Dp a^2\sda}d(M -\omega^{\prime}) \nn \\
&&\qquad -\frac{(Q -q^{\prime})r(r^2 +a^2)}{(r^2 +a^2)^2 -\Delta^{\prime}a^2\sda}d(Q -q^{\prime})\Bigg]dr \, ,
\label{KNA}
\eea
where
\bea
\Dp =  r^2 +a^2 -2(M -\omega^{\prime})r +(Q -q^{\prime})^2 = (r -r_+^{\prime})(r -r_-^{\prime}) \, , \qquad
r_{\pm}^{\prime} = M -\omega^{\prime} \pm \sqrt{(M -\omega^{\prime})^2 -(Q -q^{\prime})^2 -a^2} \, . \nn
\eea
The above integral can be evaluated by deforming the contour around the single pole $r = r_+^{\prime}$ at the
event horizon. Doing the $r$ integral first, we find
\be
\trm~ S = -\pi \int_{(M, ~Q)}^{(M -\omega, ~Q -q)} \frac{2r_+^{\prime}}{r_+^{\prime} -r_-^{\prime}}
\Big[r_+^{\prime} d(M -\omega^{\prime}) -(Q -q^{\prime}) d(Q -q^{\prime})\Big] \, .
\ee
\end{widetext}
By means of the identity
\be
(r_+^{\prime} -r_-^{\prime})dr_+^{\prime} = 2r_+^{\prime}d(M -\omega^{\prime})
-2(Q -q^{\prime})d(Q -q^{\prime}) \, ,
\ee
we can easily finish the integration and arrive at a simple expression
\be
\trm~ S = -\pi\int_{r_i}^{r_f}r_+^{\prime}dr_+^{\prime} = \frac{\pi}{2}\big(r_i^2 -r_f^2\big) \, ,
\ee
from which the tunneling rate can be readily deduced
\be
\Gamma \sim e^{-2\trm~ S} = e^{\pi(r_f^2 -r_i^2)} = e^{\Delta S_{BH}} \, ,
\ee
where $\Delta S_{BH} = S_{BH}(M -\omega, Q -q) -S_{BH}(M, Q) = \pi(r_f^2 -r_i^2)$ is the difference of
Bekenstein-Hawking entropies of the Kerr-Newman black hole before and after the particle emission. We
see that the above result perfectly generalizes those obtained in Refs. \cite{KKW,PW}, which is indicative
of a consistence with an underling unitary theory.

To end this section, it is necessary to reveal that the reason why the Kraus-Parikh-Wilczek's semi-classical
tunneling formalism is so successful is in that its effectiveness, to a large extent, relies on the well-known
thermodynamic properties of a charged rotating black hole. A rigorous check on the prescribed method allows
us to confirm that it is closely related to the first law of black hole thermodynamics. In fact, it is easily
observed that all the radial trajectories whether or not they are charged, share the common near-horizon behavior
\be
\dot{r} \approx \kappa^{\prime} (r -r_+^{\prime}) \, , \qquad
\kappa^{\prime} = \frac{r_+^{\prime} -r_-^{\prime}}{2(r_+^{\prime 2} +a^2)} \, ,
\ee
where $\kappa^{\prime}$ is the surface gravity after the particle's emission. On the other hand, note also that
the explicit expressions for the angular velocity and electric potential are given by
\bea
\Omega_+^{\prime} = \frac{a}{r_+^{\prime 2} +a^2} \, , \qquad
\Phi_+^{\prime} = \frac{(Q -q^{\prime})r_+^{\prime}}{r_+^{\prime 2} +a^2} \, , \nn
\eea
one can verify that the entropy $S^{\prime} = \pi(r_+^{\prime 2} +a^2)$ obeys the differential form of the
first law of thermodynamics \cite{WC,WSQ}
\be
d(M -\omega^{\prime}) = \frac{\kappa^{\prime}}{2\pi}dS^{\prime} +a\Omega_+^{\prime}d(M -\omega^{\prime})
+\Phi_+^{\prime}d(Q -q^{\prime}) \, .
\label{DFL}
\ee
Keeping the mass $M$ and electric charge $Q$ fixed, Eq. (\ref{DFL}) indicates that the black hole could be in
thermal equilibrium with the radiation outside the hole, and the detailed equilibrium condition is essentially
equivalent to the conservation laws established elsewhere \cite{WC}.

Substituting Eq. (\ref{DFL}) into (\ref{KNA}), then the imaginary part of the action can be rewritten as
\bea
\trm~ S &\approx& \trm \int_{r_i}^{r_f}\int_{(M, ~Q)}^{(M -\omega, ~Q -q)}
\frac{1}{\kappa^{\prime}(r -r_+^{\prime})}\Big[d(M -\omega^{\prime})  \nn \\
&&~ -a\Omega_+^{\prime}d(M -\omega^{\prime}) -\Phi_+^{\prime}d(Q -q^{\prime})\Big] dr \nn \\
&=& -\frac{1}{2} \int_{S_{BH}(M, ~Q)}^{S_{BH}(M -\omega, ~Q -q)} dS^{\prime}
= -\frac{1}{2}\Delta S_{BH} \, ,
\eea
which is equal to half of the difference of the initial and final entropy of the system. This completes our proof.

\section{Concluding remarks}
\label{CoRe}

In this paper, we have presented a neat extension of the semi-classical tunneling framework \cite{KKW,PW} in
the spherically symmetric black hole cases to deal with Hawking radiation of massless particles as a tunneling
process through the event horizon of a Kerr black hole and that of charged particles from a Kerr-Newman black
hole. The new form of the Kerr solution found by Doran \cite{NFK} is especially appropriate for us to do an
explicit tunneling calculation when transformed into the dragged Painlev\'{e}-Kerr coordinate system. By treating
the background spacetime as dynamical, the energy conservation and the angular momentum conservation as well as
the electric charge conservation are enforced in a nature way, when the particle's self-gravitation is taken into
account. Adapting this tunneling picture, we were able to compute the tunneling rate and the radiant spectrum of a
Kerr and Kerr-Newman black hole. The resulting probability of particle emission is proportional to a phase space
factor depending on the initial and final entropy of the system which multiplies the square of the quantum mechanical
tunneling amplitude for the process. Meanwhile, this implies that the emission spectrum actually deviates from
perfect thermality, but is in agreement with an underlying unitary theory (which is presumably String Theory but
that is beyond the scope of this paper).

Before concluding, some remarks are in order. First, our analysis made here strictly followed the approach \cite{KKW,PW}
that visualizes the source of Hawking radiation as a tunneling process. The pertinent point of this approach is that
black hole radiance is a dynamical mechanism for which conservation laws must be enforced. The effectiveness of the
prescribed method is closely related to the first law of black hole thermodynamics. For a stationary radiation process
where a black hole is in thermal equilibrium with the outside radiation, the detailed equilibrium condition suggests
quantum conservation laws hold true \cite{WC}. If the evaporation can stabilize with the end point of the system being
a stable remnant in thermal equilibrium with radiation, the information could be preserved. Second, we would like to
stress that the preceding study is still a semi-classical analysis (formally analogous to a WKB approximation), which
means that the radiation should be treated as point particles. Such an approximation can only be valid in the low energy
regime. If we are to properly address the information loss problem, then a better understanding of physics at the Planck
scale is a necessary prerequisite, especially that of the last stages or the endpoint of Hawking evaporation.

Finally, in a separate work \cite{WJ} the extension made here has satisfactorily been examined in the case of a
($2+1$)-dimensional rotating black hole. Further application to the case of rotating black holes in higher dimensions
is in progress.

\smallskip
{\bf Note added}: After we finished this work, the paper \cite{ZZ} appeared, discussing the charged particles'
tunnelling from the Kerr-Newman black hole.

\smallskip
{\bf Acknowledgments}: ~S.-Q.Wu was supported by a starting fund from Central China Normal University. X.Cai was
supported in part by a grant from the Natural Science Foundation of China.

\end{document}